\documentclass[aps,prd,groupedaddress]{revtex4}

\usepackage{amsfonts}
\usepackage{graphicx}

\begin{document}


\title{Mass gap in a Yang-Mills theory in the strong coupling limit}


\author{Marco Frasca}
\email[]{marcofrasca@mclink.it}
\affiliation{Via Erasmo Gattamelata, 3 \\ 00176 Roma (Italy)}


\date{\today}

\begin{abstract}
We prove that in the limit of the coupling going to infinity a Yang-Mills theory is
equivalent to a $\lambda\phi^4$ theory with the dynamics ruled just by a homogeneous
equation. This gives explicitly the Green function and the mass spectrum proving
that such gauge theories are confining. The scalar glueball spectrum is then proven
to be in fair agreement with lattice QCD computations but giving a different
ground state coinciding with the $f_0(600)$ light unflavored meson.
\end{abstract}

\pacs{11.15.Me}

\maketitle


The great success of gauge field theory to describe particle interactions has opened
up a lot of problems. Mostly of them rely on the impossibility to analyze a quantum
field theory in the limit of a very large coupling. This is the case for example of
quantum chromodynamics (QCD) where some nonperturbative results are known and perturbation
theory is reliable only at large momentum.

It is become demanding to find other approaches to perform this kind of analysis that can
grant some analytical insights into the solution of these field theories without
anyhow diminishing the relevance of a numerical approach.

The great breakthrough that started all the business of quantum gauge field theories is
due to Yang and Mills \cite{ym} that generalized the well-known U(1) symmetry gauge theory
of electromagnetism to non abelian groups. 
But a successful application of these idea had to wait the
formulation of the standard model \cite{wei,sal,gla}, the successive discovery of
asymptotic freedom \cite{gw,pol} and the proof of renormalizability by Veltman and 't Hooft
\cite{velt1,velt2,velt3}. Difficulties enter in the analysis of quantum chromodynamics
because of the strength of the low energy coupling constant that impede whatever straightforward
perturbative approach.

Recently, we introduced the duality principle in perturbation theory \cite{fra1,fra2} that permits,
by changing the choice of the perturbation terms, to get perturbative series with the development
parameter inverse each other, permitting in this way to obtain, besides the standard series thatù
holds in the limit of the coupling going to zero, also its dual series that holds in the opposite limit
of the coupling going to infinity. Rather interestingly, the leading order produces a homogeneous
equation when one considers partial differential equations.

On this way we have built a strongly coupled quantum field theory for a $\lambda\phi^4$ model \cite{fra3,fra4}
(here and in the following we take $\hbar=c=1$) with Hamiltonian
\begin{equation}
    H = \int d^{D-1}x\left[\frac{1}{2}\pi^2+\frac{1}{2}(\partial_x\phi)^2+
	\frac{1}{2}\mu_0^2\phi^2+
	\frac{1}{4}\lambda\phi^4\right].
\end{equation}
The Hamilton equations, with the space and time variables scaled through the bare mass $\mu_0$ 
as also for $\lambda$, are
\begin{eqnarray}
\label{eq:phi}
    \partial_t\phi &=& \pi \\ \nonumber
	\partial_t\pi  &=& \partial_x^2\phi -\phi -\lambda\phi^3.
\end{eqnarray}
We take
\begin{eqnarray}
   \tau &=& \sqrt{\lambda}t \\ \nonumber
   \pi &=& \sqrt{\lambda}\left(\pi_0 + \frac{1}{\lambda}\pi_1 + \frac{1}{\lambda^2}\pi_2 + \ldots\right) \\ \nonumber
   \phi &=& \phi_0 + \frac{1}{\lambda}\phi_1 + \frac{1}{\lambda^2}\phi_2 + \ldots.
\end{eqnarray}
obtaining the set of dual perturbation equations
\begin{eqnarray}
    \partial_{\tau}\phi_0 &=& \pi_0 \\ \nonumber
	\partial_{\tau}\phi_1 &=& \pi_1 \\ \nonumber
    \partial_{\tau}\phi_2 &=& \pi_2 \\ \nonumber
	                 &\vdots&  \\ \nonumber
	\partial_{\tau}\pi_0 &=& -\phi_0^3 \\ \nonumber
	\partial_{\tau}\pi_1 &=& -3\phi_0^2\phi_1 + \partial_x^2\phi_0-\phi_0 \\ \nonumber
	\partial_{\tau}\pi_2 &=& -3\phi_0\phi_1^2-3\phi_0^2\phi_2 + \partial_x^2\phi_1-\phi_1\\ \nonumber
	                 &\vdots&
\end{eqnarray}
and one sees that at the leading order a homogeneous equation rules the dynamics of the field.
Then, one can prove numerically that in the limit $\lambda\rightarrow\infty$ for the classical theory
one can still use a Green function method \cite{fra4} with the leading order homogeneous equation
\begin{equation}
    \ddot G+\lambda G^3=\delta(t),
\end{equation}
having restated expicitly $\lambda$. This equation has the exact solution
\begin{equation}
\label{eq:gf}
    G(t)=\theta(t)\left(\frac{2}{\lambda}\right)^{\frac{1}{4}}
	{\rm sn}\left[\left(\frac{\lambda}{2}\right)^{\frac{1}{4}}t,i\right].
\end{equation}
and its time reversed version, with $\rm sn$ the snoidal elliptic Jacobi function. 
The quantum field theory is then given by
\begin{equation}
    Z[j]=\exp\left[\frac{i}{2}\int d^Dy_1d^Dy_2\frac{\delta}{\delta j(y_1)}(-\nabla^2+1)\delta^D(y_1-y_2)
    \frac{\delta}{\delta j(y_2)}\right]Z_0[j]
\end{equation}
being
\begin{equation}
    Z_0[j]=\exp\left[\frac{i}{2}\int d^Dx_1d^Dx_2j(x_1)\Delta(x_1-x_2)j(x_2)\right]
\end{equation}
with the Feynman propagator
\begin{equation}
    \Delta(x_2-x_1)=\delta^{D-1}(x_2-x_1)[G(t_2-t_1)+G(t_1-t_2)].
\end{equation}
Noting the following relation for the snoidal function\cite{gr}
\begin{equation}
    {\rm sn}(u,i)=\frac{2\pi}{K(i)}\sum_{n=0}^\infty\frac{(-1)^ne^{-(n+\frac{1}{2})\pi}}{1+e^{-(2n+1)\pi}}
    \sin\left[(2n+1)\frac{\pi u}{2K(i)}\right]
\end{equation}
being $K(i)$ the constant
\begin{equation}
    K(i)=\int_0^{\frac{\pi}{2}}\frac{d\theta}{\sqrt{1+\sin^2\theta}}\approx 1.3111028777,
\end{equation}
the Fourier transform of the Feynman propagator is
\begin{equation}
    \Delta(\omega)=\sum_{n=0}^\infty\frac{B_n}{\omega^2-\omega_n^2+i\epsilon}
\end{equation}
being
\begin{equation}
    B_n=(2n+1)\frac{\pi^2}{K^2(i)}\frac{(-1)^{n+1}e^{-(n+\frac{1}{2})\pi}}{1+e^{-(2n+1)\pi}},
\end{equation}
and the mass spectrum of the theory given by
\begin{equation}
    \omega_n = \left(n+\frac{1}{2}\right)\frac{\pi}{K(i)}\left(\frac{\lambda}{2}\right)^{\frac{1}{4}}
\end{equation}
proper to a harmonic oscillator. In this case, the mass gap in the limit $\lambda\rightarrow\infty$ is
\begin{equation}
\label{eq:ds}
     \delta_S = \frac{\pi}{2K(i)}\left(\frac{\lambda}{2}\right)^{\frac{1}{4}}
\end{equation}
corresponding to the choice $n=0$, that is produced by the self-interaction of the scalar field. As shown
in Ref.\cite{fra4}, the next correction in the dual perturbation series is proportional to the
missing term in the propagator, i.e. ${\bf k}^2+1$ as it should be.

Our aim is to prove that a similar result does hold for a Yang-Mills gauge theory in the limit of the
coupling constant going to infinity for the gauge group SU(N). The Hamilton equations 
in the gauge $A_0^a=0$ can be written down as \cite{fs,smi}
\begin{eqnarray}
    \partial_t A_k^a&=&F_{0k}^a \\ \nonumber
    \partial_t F_{0k}^a&=&\partial_lF_{lk}^a+gf^{abc}A_l^bF_{lk}^c
\end{eqnarray}
being $g$ the coupling constant, $f^{abc}$ the structure constants of the gauge group,
$F_{lk}^a=\partial_lA_k^a-\partial_kA_l^a+gf^{abc}A_l^bA_k^c$ and the constraint
$\partial_kF_{0k}^a+gf^{abc}A_k^bF_{0k}^c=0$ does hold. So, let us introduce the
following equations, as done for the scalar field,
\begin{eqnarray}
   \tau &=& gt \\ \nonumber
   F_{0k}^a&=& gF_{0k}^{a(0)} + F_{0k}^{a(1)} + \frac{1}{g}F_{0k}^{a(2)} + \ldots \\ \nonumber
   F_{lk}^a&=& F_{lk}^{a(0)} + \frac{1}{g}F_{lk}^{a(1)} + \frac{1}{g^2}F_{lk}^{a(2)} + \ldots \\ \nonumber
   A_k^a &=& A_k^{a(0)} + \frac{1}{g}A_k^{a(1)} + \frac{1}{g^2}A_k^{a(2)} + \ldots.
\end{eqnarray}
Being $g$ adimensional, we now suppose to have scaled space and time variables with an arbitrary energy
scale $\mu_0$ in order to make the problem similar to that of the scalar field that has a natural energy scale
in the bare mass. This scale arises naturally from the gauge invariance and appears
independently from the lattice spacing in lattice QCD otherwise in the continuum limit
one would get a zero mass gap.  So, at the end of the computation we will get an adimensional mass gap. 
Finally, one has the perturbation equations
\begin{eqnarray}
    \partial_\tau A_k^{a(0)}&=&F_{0k}^{a(0)} \\ \nonumber
    \partial_\tau A_k^{a(1)}&=&F_{0k}^{a(1)} \\ \nonumber
    &\vdots& \\ \nonumber
    \partial_\tau F_{0k}^{a(0)}&=&f^{abc}f^{cde}A_l^{b(0)}A_l^{d(0)}A_k^{e(0)} \\ \nonumber
    \partial_\tau F_{0k}^{a(1)}&=&
    f^{abc}f^{cde}A_l^{b(1)}A_l^{d(0)}A_k^{e(0)}
    +f^{abc}f^{cde}A_l^{b(0)}A_l^{d(1)}A_k^{e(0)}
    +f^{abc}f^{cde}A_l^{b(0)}A_l^{d(0)}A_k^{e(1)} \\ \nonumber
    & &+f^{abc}\partial_l\left(A_l^{b(0)}A_k^{c(0)}\right)
    +f^{abc}A_l^{b(0)}\left(\partial_lA_k^{c(0)}-\partial_kA_l^{c(0)}\right) \\ \nonumber
    &\vdots&
\end{eqnarray}
and we can recognize at the leading order the homogeneuos Yang-Mills equations as promised. These equations
display a rich dynamics as e.g. Hamiltonian chaos \cite{sav1,sav2,sav3} and can be derived from the following
Hamiltonian
\begin{equation}
     H_{YM} = \frac{1}{2}\dot A_l^{a(0)}\dot A_l^{a(0)}+
     \frac{1}{4}f^{abc}f^{ade}A_l^{b(0)}A_k^{c(0)}A_l^{d(0)}A_k^{e(0)}.
\end{equation}
This Hamiltonian becomes the same as the leading order Hamiltonian for a scalar field if the potentials
$A_l^{a(0)}$ are properly chosen as in \cite{Frasca:2007uz}. This solution is generally unstable with respect to small
perturbations but we recall here that we are considering the opposite limit of a perturbation
going to infinity as pointed out in Ref.\cite{fra4} and instability does not apply here.
In order to do this we introduce the reduced
't Hooft symbols\cite{th1,th2} $\eta_i^a$ so to have $\eta_i^a\eta_i^b=\delta^{ab}$ and
the solution can be written as $A_l^{a(0)}(t)=\eta_l^a A(t)$. 
This gives the Hamiltonian for a SU(N) theory
\begin{equation}
     H_{A} = \frac{1}{2}(N^2-1)\dot A^2+\frac{1}{4}N(N^2-1)A^4.
\end{equation}
where use has been made of the relation $f^{abc}f^{abc}=N(N^2-1).$
This is the only Hamiltonian that gives a stable quantum field theory in the limit $g\rightarrow\infty$
mapping the Yang-Mills gauge theory, in the infrared limit, to a $\lambda\phi^4$ theory
that has been proved to have an infrared perturbation theory \cite{fra4}.
This can be seen very easily by taking one of the potentials as $A(t)+\epsilon(t)$ being $\epsilon(t)$ a
small variation. The term that is so gained by the Hamiltonian is not finite in the limit $g\rightarrow\infty$
producing an infinite energy with respect to our case. So, we are granted to have a stable quantum field theory
in the limit of a large coupling only by taking at the leading order properly chosen potentials \cite{Frasca:2007uz}. 
We emphasize that this argument is crucial in order to reach our aim and should be considered
as a possible evidence that to build a strongly coupled quantum field theory one needs an integrable
system. The only way to achieve this for a Yang-Mills field theory is to solve for $A(t)$.
Then, a SU(N) Yang-Mills theory is ruled at the leading order by the following nonlinear equation 
after the time $t$ has been reinserted
\begin{equation}
     \ddot A + g^2NA^3=0
\end{equation}
as for the $\lambda\phi^4$ theory. The mass gap is immediately read out from eq.(\ref{eq:ds})
taking $\lambda=g^2N$ and is given by
\begin{equation}
\label{eq:dm}
     \delta_{YM} = \frac{\pi}{2K(i)}\left(\frac{g^2N}{2}\right)^{\frac{1}{4}}
\end{equation}
proving that the self-interaction of a Yang-Mills field is confining in the limit of a large coupling.
We note that also in the strong coupling limit the 't Hooft scaling $g^2N$ does hold
as also shown in lattice Yang-Mills quantum field theory computations \cite{tep}. 
We can also write out a leading order mass spectrum reading it from the scalar field as 
$\omega_n=\left(n+\frac{1}{2}\right)\frac{\pi}{K(i)}\left(\frac{g^2N}{2}\right)^{\frac{1}{4}}$.
As for the scalar field we expect that the terms with the momentum $\bf k$ should appear in higher order
corrections to the propagator in agreement with the K\"allen-Lehman representation and as has been
proved for the scalar field \cite{fra4}. We also point out that the mass dimensionality arises from
the arbitrary energy scale we have introduced to scale time and space variables. 
In order to complete the comparison with lattice QCD we firstly point out that a recent
result on the gluon propagator confirms that this is independent on the position in the
infrared regime \cite{lm} in agreement with our approach. Finally we give the glueball
spectrum to be compared with \cite{tep2,morn}. We get from eq.(\ref{eq:dm}) for the scalar
glueball and its excited states
\begin{equation}
    \frac{m_G}{\sqrt{\sigma}}=1.198140235(2n+1)
\end{equation}
where we have put the tension $\sigma=\sqrt{g^2N/2}\mu_0^2$ being
$\mu_0$ the scale that in lattice QCD is chosen to give \cite{morn} $\sqrt{\sigma}=410(20)$ MeV. So,
one has a lower state in QCD for n=0 with mass $m_G=491(20)$ Mev that hits quite
well the light unflavored meson $f_0(600)$ \cite{par}. We can also consider the same computation
by taking \cite{tep0} $\sqrt{\sigma}=440(38)$ MeV that gives  $m_G=527(38)$ in fully
agreement with the mass $m_G=528(32)$ obtained in Ref.\cite{gal}.
This lower state cannot be seen in the current
lattice computations due to the coarse graining introduced by the lattice spacing. For
$n=1$, $n=2$ and $n=3$ one has
\begin{eqnarray}
    \frac{m_G}{\sqrt{\sigma}}(n=1)&=&3.594420705 \\ \nonumber
	\frac{m_G}{\sqrt{\sigma}}(n=2)&=&5.990701175 \\ \nonumber
	\frac{m_G}{\sqrt{\sigma}}(n=3)&=&8.386981645
\end{eqnarray} 
in very good agreement with the values in ref.\cite{tep2} $3.55(7)$ and $5.69(10)$ for
SU(3) and $3.307(53)$ and $6.07(17)$ for the extrapolated values for $N$ to infinity. The case
$n=3$ is our prevision for the next excited state of the scalar glueball to be found in
lattice calculations. 

In conclusion we have proved that in the limit of a coupling constant going to infinity a SU(N)
Yang-Mills gauge field produce a mass gap confirming its confining property being the dynamics
ruled by a similar equation as for a scalar self-interacting field. A strongly coupled quantum
field theory can be possibly built on these bases.


\end{document}